%% file: nimo.tex
\input{macros1.tex}

\input{macros2.tex}
\documentstyle[12pt,psfig]{ioplppt}
\begin{document}
\title {\bf Magnetic properties of Ni-Mo single crystal alloys : theory and experiment} 
\author{ Subhradip Ghosh\ftnote{1}{Corresponding author. E-mail : subhra@boson.bose.res.in} and Nityananda Das} 
\address{S.N. Bose National Centre for Basic Sciences,
 JD Block, Sector 3, Salt Lake City, Calcutta 700091,
India.}
\author{Abhijit Mookerjee\ftnote{3}{On sabbatical leave from : S.N. Bose National Centre for Basic
Sciences, India}}
\address{Department of Physics, Indian Institute of Technology,Kanpur 208016, India}
\begin{abstract}
The magnetization of Ni$_{1-x}$Mo$_{x}$ single crystals with x=4,6,8 and
10 $\%$ by weight have been measured at 4.2K using a vibrating sample
magnetometer and a Superconducting Quantum Interference Device (SQID).
The magnetization of the alloy at these low concentrations and at 0K
have been theoretically determined by using the tight-binding lnearized
muffin-tin orbital method coupled with augmented space recursion. The
theoretical data are compared wih the experiment.
\end{abstract}

\section{Introduction}
The equilibrium phase diagram of the Ni$_{1-x}$Mo$_{x}$
 system exhibits a continuous face centred cubic disordered solid solution
 in the range 0$<$x$\le$ 0.12. The alloy shows a series of
 ordered phases at higher concentrations \cite{kn:mass}.
 Das \etal
 \cite{kn:gpd} have 
 studied the phase diagram at the higher concentration range using
 the local density approximation (LDA) based tight-binding linearized
 muffin-tin approximation (TB-LMTO) method proposed by Andersen \etal
 \cite{kn:ajs}. To our  knowledge a similar study of the disordered
 regime has not been carried out in detail. In this communication we shall report
 such a study using the same TB-LMTO methodolgy as before 
 but combining this with
 the augmented space recursion proposed by one of us \cite{kn:mook} 
 to take care of the disorder configuration averaging. We shall evaluate
 the local magnetization as a function of the Mo concentration. Simultaneously,
 we shall measure the magnetization of four alloy systems in this
 disordered alloy regime (x=0.04, 0.06, 0.08 and 0.1). Side by side we shall
compare our results with  experimental work on the magnetization and Curie temperatures
of single crystals of NiMo \cite{kn:kan}.

 \section{Theoretical Details}

 The TB-LMTO has been described in great detail  earlier \cite{kn:akm}
 . We shall refer the reader to the referenced monograph for technical
 details.

Description of  magnetic phases within the local spin density approximation
(LSDA) involves the study
of the evolution of local magnetic moments in the vicinity of ion
cores because of the distribution of the valence electron charge.
Each lattice site in the face centred cubic structure is occupied
by an ion core : in our case randomly by either Ni or Mo. We shall
 associate a cell or a sphere  with each ion core  and assume that the charge
 contained in the sphere belongs to that ion
core alone. Ideally such cells or spheres should not overlap.
In the traditional Kohn-Korringa-Rostocker (KKR) method this
is certainly so. However, in the atomic sphere approximation (ASA) 
which we shall use in our TB-LMTO version, this  
division of space is to a certain extent arbitrary.  Within these
cells the valence electrons carrying spin $\sigma$ sees a binary
random spin-dependent potential V$^{\lambda}_{\sigma}$($\underline{r}$),
where $\lambda$\equal Ni or Mo and $\sigma$\equal $\uparrow$ or $\downarrow$.

The charge density within the cells can be obtained from solving the
Schr\"odinger equation within the LSDA. The charge density over
the solid can be written as :

\[
\fl \rho_{\sigma}(\ul{r})\equal -(1/\pi)\Im m \sum_{L} \int_{-\infty}^{E_{F}}
\left[ x \ll G^{Mo,\sigma}_{LL}(\ul{r},\ul{r},E)\gg 
+ (1-x) \ll G^{Ni,\sigma}_{LL}(\ul{r},\ul{r},E)\gg
\right] dE
\]

where,
\vskip 0.05cm

$\ll G^{Mo,\sigma}_{LL}(\ul{r},\ul{r},E)\gg$  and 
$\ll G^{Ni,\sigma}_{LL}(\ul{r},\ul{r},E)\gg$ are partially
averaged Green functions with the site $\ul{r}$ occupied  by a 
Mo or Ni ion core potential corresponding to spin $\sigma$. The
Mo sites are almost spin independent (except for a very small induced
moment) and do not appreciably contribute to local moment densities. 

The averaging is done over configurations of the random alloy.
 A powerful technique of carrying out this averaging is the augmented space recursion
\cite{kn:asr}. The method allows us to go well beyond the traditional
single site coherent potential approximations and has been applied
successfully to a wide variety of systems \cite{kn:pm,kn:dm}. The convergence
of the ASR has been established recently \cite{kn:gdm}, so that any
approximation we impose on the recursion is controlled by tolerence
limits preset by us.

For the random ferromagnetic phase we proceed as follows : we consider all
cells to be identical in that they all carry identical average charge densities.
We shall borrow the notation of Andersen \etal\cite{kn:ajs} to write
functions like $\~f(\ul{r}_{R}$) which are equal to $f(\ul{r})$ when $\ul{r}$ lies
in the atomic sphere labelled by R and is zero outside. The ferromagnetic charge densities
are defined as :

\begin{eqnarray*}
\rho_{1}(\ul{r})\equal \sum_{R} \~\rho_{\uparrow}(\ul{r}_{R}) 
 \\
\rho_{2}(\ul{r})\equal \sum_{R} \~\rho_{\downarrow}(\ul{r}_{R}) 
\end{eqnarray*}

The magnetic moment  per cell (atom) is then defined by :

\begin{eqnarray*}
 m & = & (1/N) \int d^{3}\ul{r} \left[ \rho_{1}(\ul{r})\minus
\rho_{2}(\ul{r})\right] \\
   & = & (1/N) \sum_{R} \int_{r\le S} d^{3}\ul{r} \left[ \~\rho_{\uparrow}(r_{R})
\minus \~\rho_{\downarrow}(r_{R})\right] \\
& = & (1/N) \sum_{R} \int_{r\le S} d^{3}\ul{r}\enskip
 m_{R}(r_{R}) \\
\end{eqnarray*}

Since all cells are indentical, the above calculation need be done only
in one typical cell. Within the TB-LMTO-ASA the cells are replaced by
inflated atomic spheres and the remaining interstitial is neglected. The 
problem is then one of a binary alloy with an almost non-magnetic charge density
due to the Mo ion cores and a magnetic one due to the Ni ones.

For the calculation of the component projected averaged density of states
of the ferromagnetic
 phase we have used a real space cluster of 400 atoms and
an augmented space shell upto the sixth nearest neighbour from the starting
state. Eight pairs of recursion coefficients were determined exactly and the
continued fraction terminated by the analytic terminator due to Luchini
and Nex \cite{kn:ln}. In a recent paper Ghosh \etal \cite{kn:gdm} have shown the
convergence of the related integrated quantities,  like the Fermi energy,
the band energy, the magnetic moments and the charge densities, within the augmented space
recursion. The convergence tests suggested by the authors were carried
out to prescribed accuracies. We have reduced the computational burden of the
recursion in the full augmented space by using the local symmetries of the
augmented space to reduce the effective rank of the invariant subspace
in which the recursion is confined \cite{kn:asr} and using the seed
recursion methodology \cite{kn:gm} with fifteen energy seed points
uniformly across the spectrum. Both the reduction techniques have been
described in detail in the referenced papers and the readers are
referred to them for details. It is important to emphasize this point,
since there has been erroneous statements made earlier that although the augmented
space recursion method is attractive mathematically, it was not feasible
for application as a computational technique is realistic alloys. Further,
it has been shown \cite{kn:asr}  that augmented space recursion with an
analytic terminator {\sl always} produces herglotz results, whether we
use the homogeneous disorder model as in this paper or the version
including short-ranged order \cite{kn:pm} or local lattice distortions
\cite {kn:dm}.

We have chosen the Wigner-Seitz radii of the two constituent atoms Mo
and Ni in such a way that the average volume occupied by the atoms is
conserved. Within this constraint we have varied the radii so that
the final configuration has neutral spheres. This eliminates the
necessity to include the averaged Madelung Energy part in the total
energy of the alloy. The definition and computation of the Madelung
Energy in a random alloy had faced controversy in recent literature
 and to this date no satisfactory resolution of the problem
exists. Simultaneously we have made sure that the sphere overlap remains
within the 15$\%$ limit prescribed by Andersen.

The calculations have been made self-consistent in the LSDA sense, that
is, at each stage the averaged charge densities are calculated from the
augmented space recursion and the new potential is generated by the
usual LSDA techniques. This self-consistency cycle was converged in
both total energy and charge to errors of the order 10$^{-5}$. We have
also minimized the total energy with respect to the lattice constant.
The quoted results are those for the minimum configuration. No short
ranged order due to chemical clustering has been taken into account
in these calculations, nor any lattice distortions due to the size
differences between the two constituents.

The estimates of the Curie temperature were obtained from the magnetic pair energies
\cite{kn:mook2}
The pair energies are defined as follows : 
At two sites labelled $\ul{r}$ and $\ul{r'}$ in a completely random paramagnetic
background, we replace the potential by that
of either the up-spin ferromagnetic Ni or the down-spin one. The Green function
of this system we shall denote by : $G_{LL}^{Ni,\sigma\sigma '}(\ul{r},\ul{r},E)$, 
$\sigma$ being the spin type
at the site $\ul{r}$ (either $\uparrow$ or $\downarrow$) and $\sigma '$ that at the site $\ul{r'}$. The
pair energy is defined as 

\begin{eqnarray*}
 E({\ul{R}}) &\eq &\int_{-\infty}^{E_{F}} dE\;E\;\left[-(1/\pi)\Im m \left(
G_{LL}^{Ni,\uparrow\uparrow}(\ul{r},\ul{r},E)+G_{LL}^{Ni,\downarrow\downarrow}
(\ul{r},\ul{r},E)\ldots\right.\right.\\
& &\left.\left. -G_{LL}^{Ni,\uparrow\downarrow}(\ul{r},\ul{r},E)-G_{LL}^{Ni,\downarrow\uparrow}(
\ul{r},\ul{r},E) \right)\right]
\end{eqnarray*} 

Here $\ul{R}\eq \ul{r}-\ul{r'}$.
We may either estimate the above directly, or to be more accurate we may use the
orbital peeling method of Burke \cite{kn:bk}. The latter is an extension of the
recursion method, where small differences of large energies (as in the definition
of the pair energy) are obtained directly and accurately from the recursion continued
fraction coefficients. Note that we have assumed that the dominant contribution
to the pair energy comes from the band contribution and the rest approximately cancel out. The
simplest Bragg-Williams estimate of the Curie temperature is 

\[ T_{c} \eq (1-x)E(\ul{0})/\kappa_{B} \]  where 

$E(\ul{0})\; =\; E(\ul{q}=\ul{0})$ and 
$E(\ul{q})\; =\;\sum_{\ul{R}} \exp{\left(i\ul{q}.\ul{R}\right)}E(\ul{R})$. Since the pair energy is short-ranged, a
reasonable estimate of $E(\ul{0})$ is $ \sum_{n<3} Z_{n}\;E(\ul{R}_{n})$ where $\ul{R}_{n}$ is the n$^{th}$-nearest

neighbour vectors and $Z_{n}$ is the number of n$^{th}$-nearest neighbours. The Bragg-Williams
approach overestimates the Curie temperature and its generalization, the
cluster variation method, yields better quantitative estimates. We have restricted
ourselves to the Bragg-Williams, nearest neighbour pair energy approximation.

\section{Results and Discussion}

In the phase diagram for NiMo alloys we notice first that
the solid solution phase occupies a small part of the phase diagram below x=0.12
and extends down upto the lowest temperatures. In this region there is no
transition to an ordered phase at low temperatures. This is the concentration
region that we have focussed on in this communication.

\begin{figure}[h]
\centering
\hfil\psfig{file=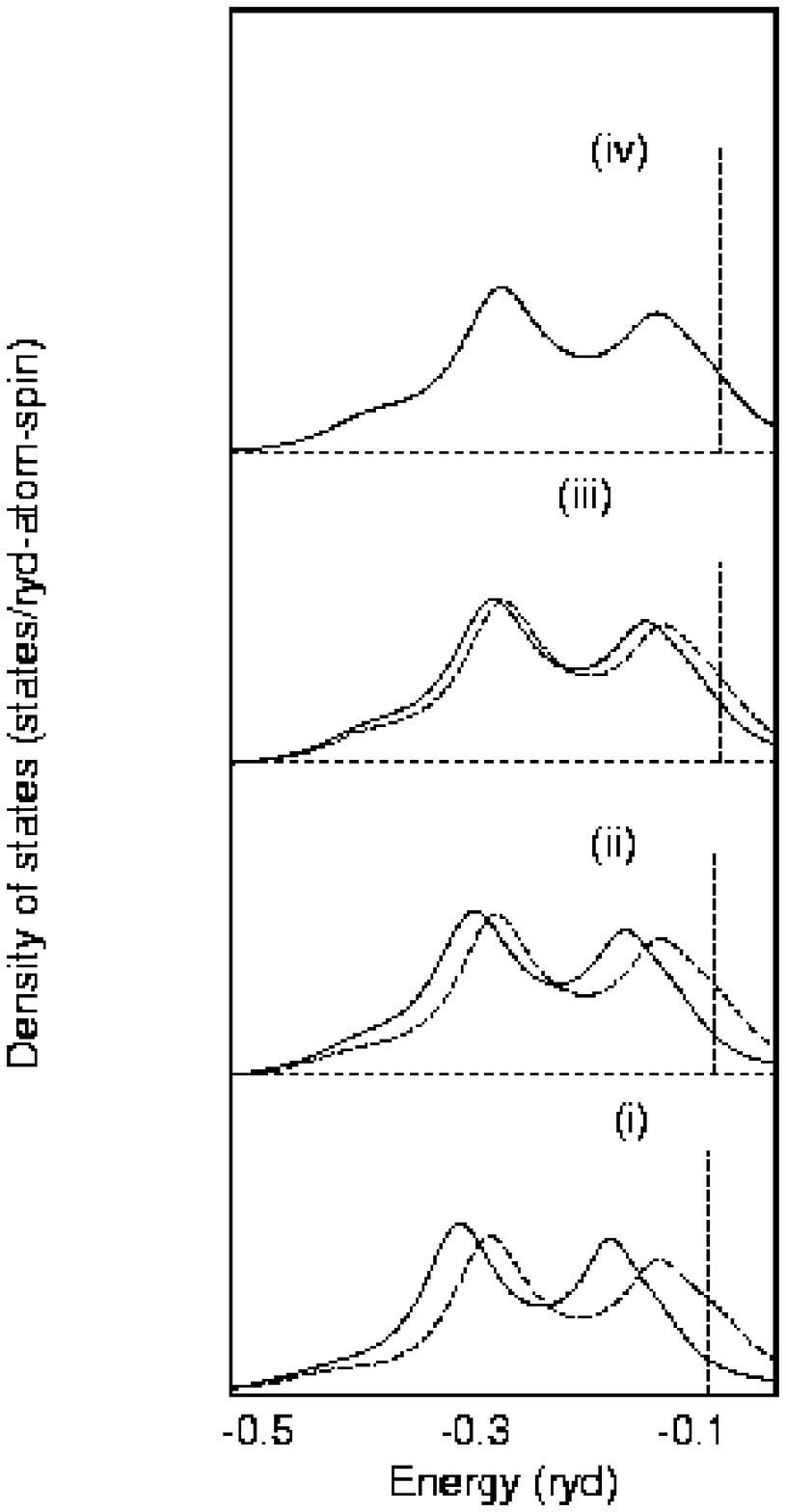,height=10cm,width=10cm}\\
\hfil\psfig{file=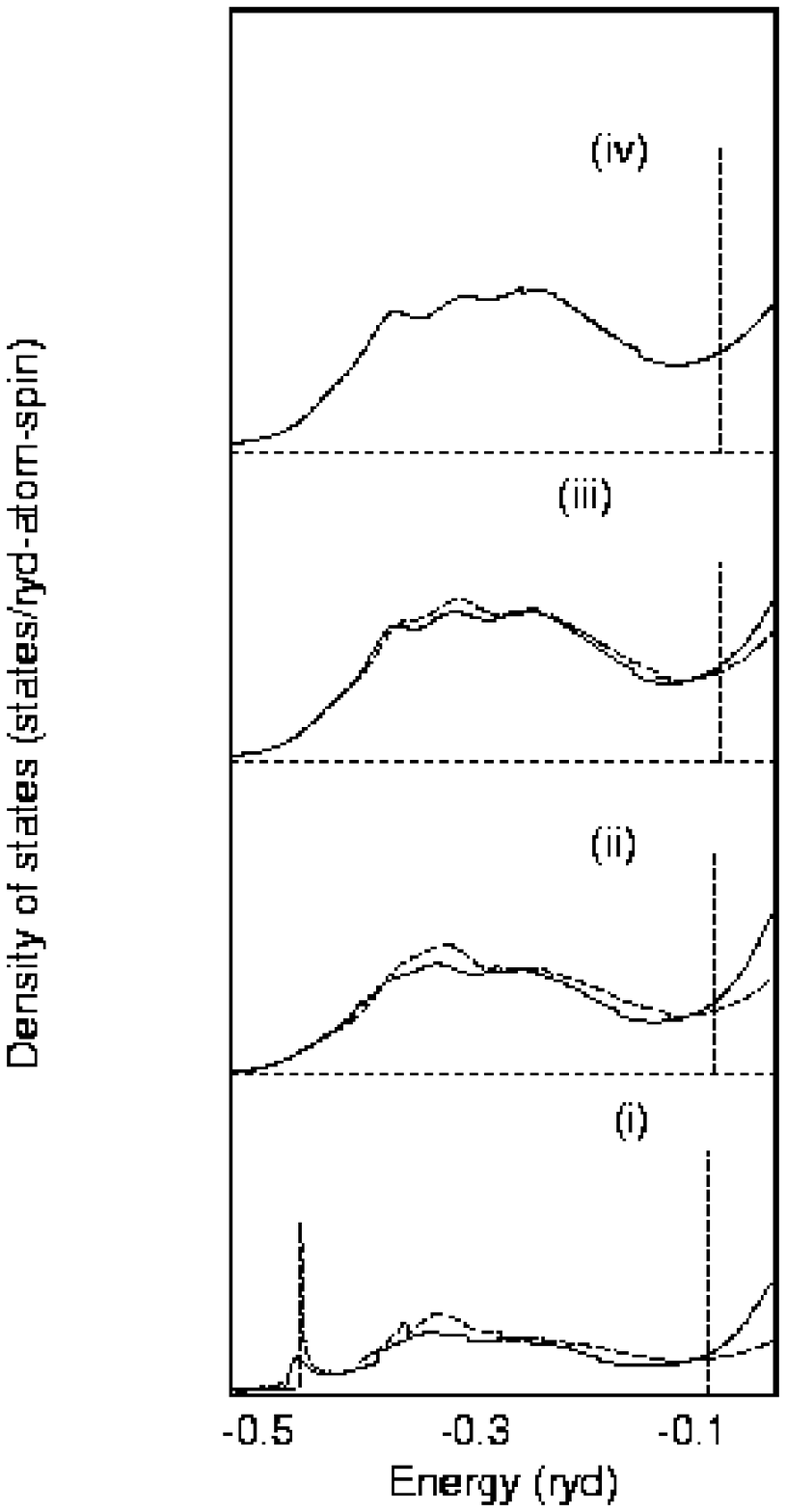,height=10cm,width=10cm}\\
\caption{ The partial densities of states of Ni and Mo
at 2,6,10 and 14 atomic $\%$ of Mo. Dashed curves show the results for
the minority and full curve the majority spin states.}
\end{figure}

 Figure 1 shows the partial density of states for Ni and Mo. The full curves show the majority spin partial density of states and the dashed curves those for the minority spin. The Mo atomic concentrations are (i) 2$\%$ (ii) 4$\%$ (iii) 10$\%$ and (iv) 14$\%$. The exchange splitting of the Ni {\it d}-states decrease with increasing Mo concentration. Exchange effects on Mo is very small and there is a very small induced moment on Mo atoms at low concentrations. Since we are interested in the shape of the density of states, the figure shows them in arbitrary units scaled between 0 and 1. The Fermi energies lie in the region just above -0.2 ryd.

\begin{figure}[h]
\centering
\hfil\psfig{file=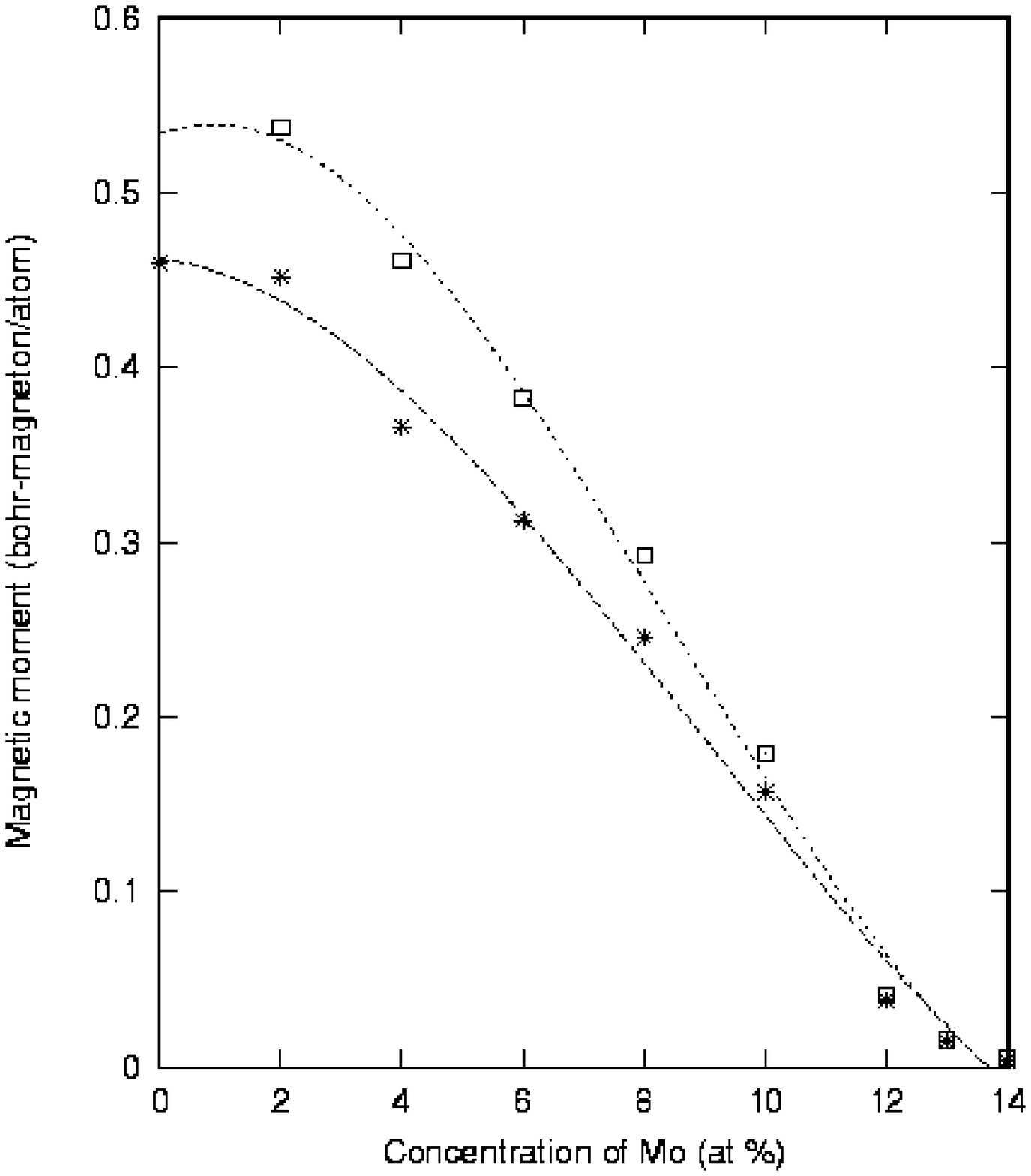,height=10cm,width=10cm}\\
\caption{The magnetic moment on Ni as a function
of Mo concentration. Stars refer to the augmented space calculations
while the squares to the CPA.}
\end{figure}

\begin{figure}[h]
\centering
\hfil\psfig{file=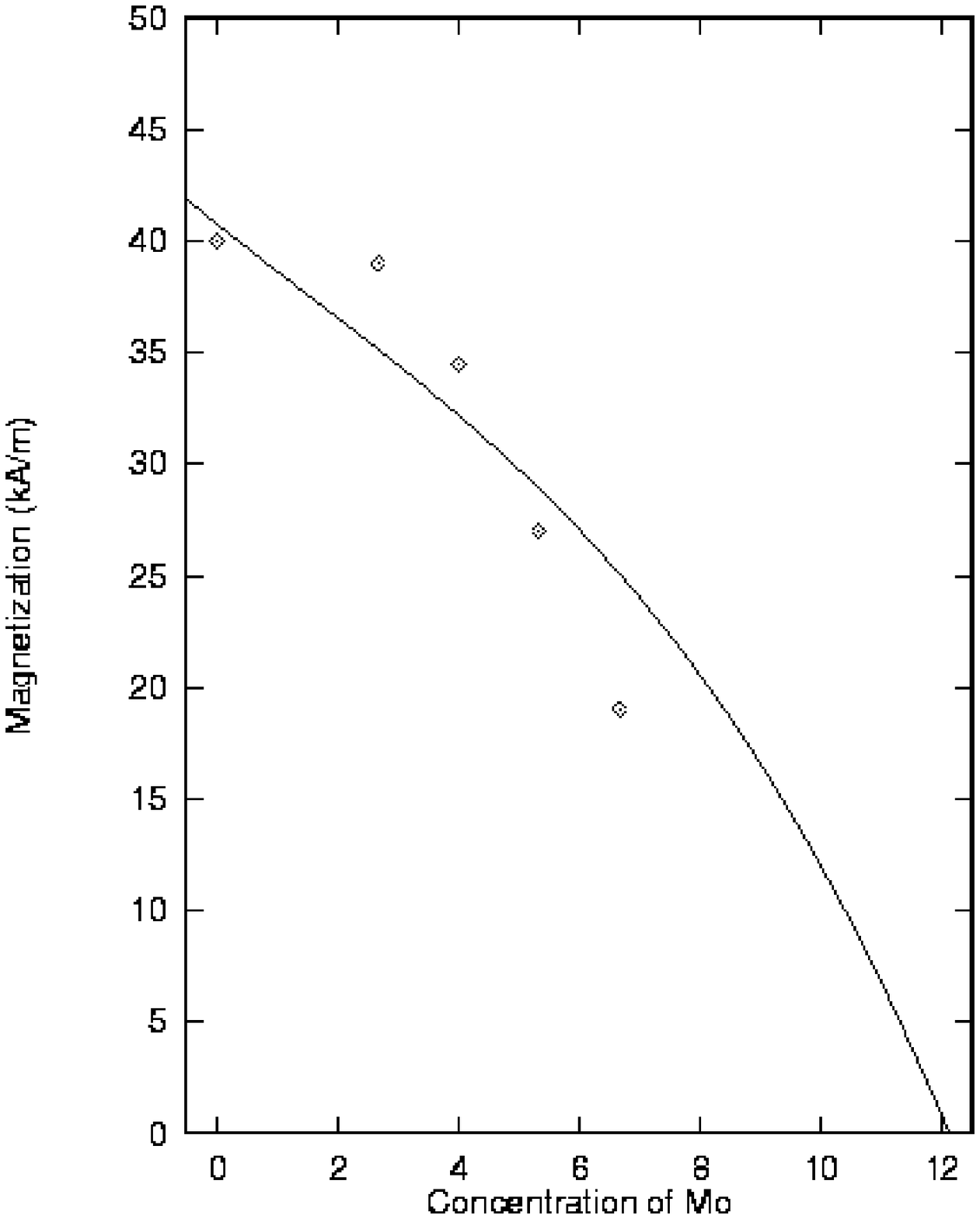,height=10cm,width=10cm}\\
\caption{The magnetization as a function of Mo concentration
at 0K obtained from the theoretical estimates.}
\end{figure}

 Figure 2 shows the local magnetic moment in bohr-magneton/atom as a function of the Mo concentration. The short dashed curve are CPA results,while the long dashed curve shows the results from the ASR. In the low concentration regime, the CPA consistently gives larger magnetic moments. To compare with experiment, we convert the magnetic monemt to magnetization in units of kA/m. The ASR results are shown in figure 3. The experimental data of Khan {\it et al} \cite{kn:kan} are shown as squares. We first note that the ASR results for the very low Mo concentration regime agree rather well with experiment, while the CPA results are consistently higher. Khan {\it et al} suggest that the magnetization vanishes around concentration of 8$\%$ of Mo. The rigid band model predict a transition around 10$\%$, while both the CPA and the ASR predict a transition around 12 to 13$\%$ of Mo. How do we reconcile these discrepancies ? The rigid band model does not take into account charge transfer due to alloying between Ni and Mo, so cannot be quantitatively accurate. Our theoretical results actually calculate the magnetization per atom in a ferromagnetic arrangement. Here the local and global magnetizations are the same. The experimental results yield the average global magnetization $m=(1/N)\sum m_{loc}$. NiMo like all typical spin-glass alloys, is a solid solution of a magnetic constituent Ni and a non-magnetic one, Mo. Therefore,
like all spin-glass alloys, we expect a paramagnetic-spin-glass transition
around the concentration region 8$\%$-13$\%$ of Mo. The experimental global
magnetization experiments will show a vanishing magnetization, whereas, our
calculations will not show this. More detailed experiments like Mossbauer, low
-field dc-susceptibility, hysterisis and so on, need to be done in this concentration region to get a better picture. This regime promises richness and variety
 in magnetic behaviour.

\begin{figure}[h]
\centering
\hfil\psfig{file=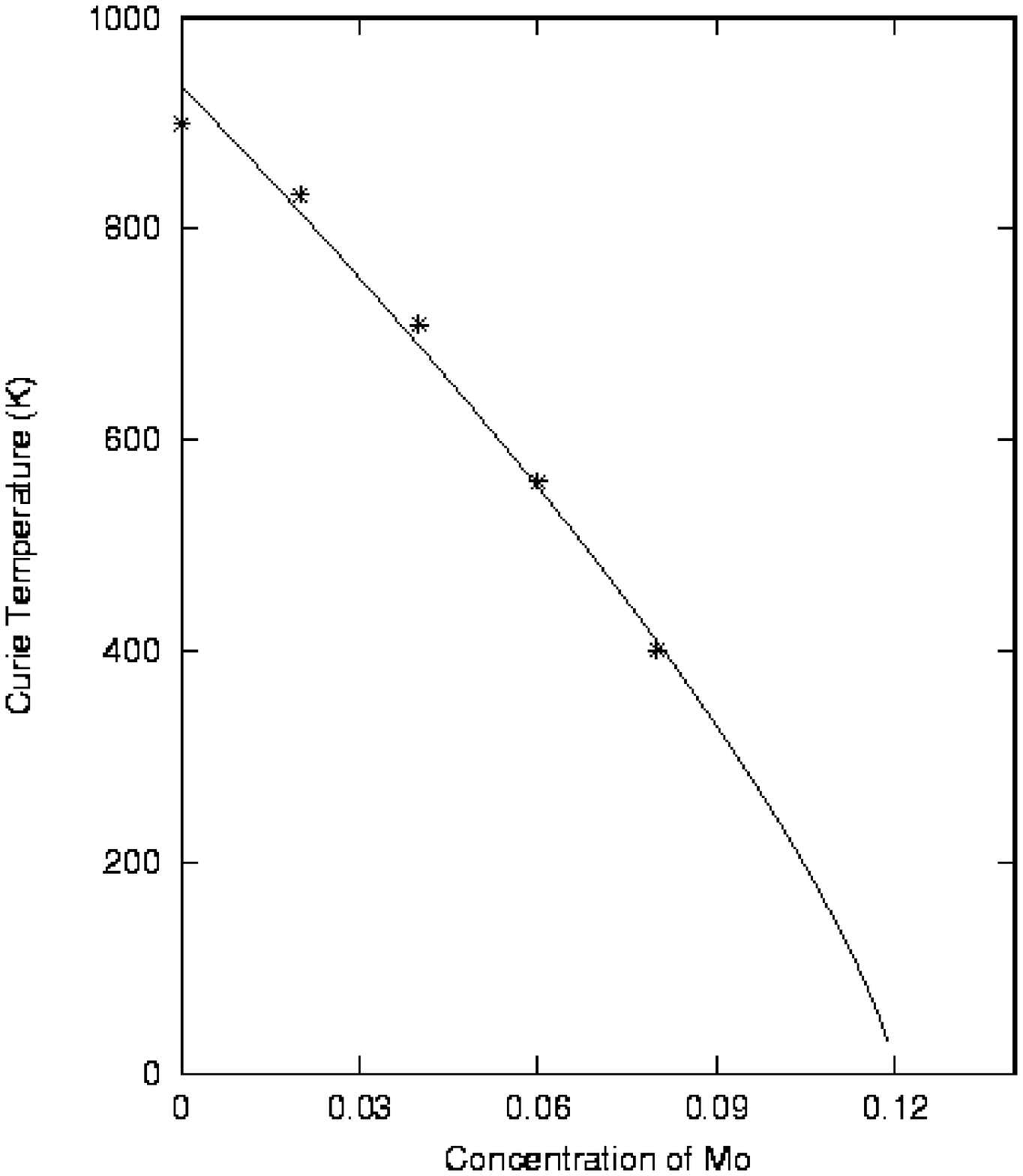,height=10cm,width=10cm}\\
\caption{The Curie temperature as a function of Mo concentration.}
\end{figure}

 Figure4 plots the Curie temperature versus concentration of Mo. Qualitatively
 the bahaviour is in agreement with the results of Khan {\it et al}. The
Bragg-Williams approximation used here is known to consistently overestimate
the transition temperature. These results also indicate absence of transition
from paramagnetic to an ordered phase at around 12$\%$ of Mo. Again, this is
not surprising in the context of the discussion above. Our theoretical model
also does not incorporate the possibility of a spin-disordered phase which
becomes energetically favourable at around 8$\%$ Mo concentration.

 We conclude with the remark that the concentration regime 8$\%$-14$\%$ Mo
requires both more careful experimental studies as well as more elaborate
theoretical models which incorporate the possibility of the spin-disordered
phase as well.

\section*{Acknowledgements}
ND would like to thank the CSIR, India for financial assitance.
AM acknowledges useful discussions with Dr. G.P. Das and Prof. A.K.
Majumdar.

\end{document}

%% file: macros1.tex
  \def\ket{\vert \vert  \{ \emptyset \} \rangle}
  \def\ket2{\vert \vert \otimes \{ R \} \rangle}

\def\.#1{\mathaccent 95#1}
\def\^#1{\mathaccent 94 #1}
\def\~#1{\mathaccent "7E #1}

\def\equal{\enskip =\enskip}

\def\minus{\enskip -\enskip}
\def\eq{\enskip =\enskip}

\def\ul#1{\underline{#1}}

  \def\ket{\vert \vert  \{ \emptyset \} \rangle}
  \def\ket2{\vert \vert \otimes \{ R \} \rangle}